\documentclass[aps,twocolumn,showpacs,superscriptaddress]{revtex4-1}

\newcommand{\mueff}{\mu^\ast}
\newcommand{\EG}{E_{\mbox{\scriptsize\it G}}}

\newcommand{\MSAT}{M_{\mbox{\tiny SAT}}}

\newcommand{\TN}{T_{\mbox{\tiny N}}}

\newcommand{\thetaF}{\theta_F}

\usepackage{epsfig,bm,amsmath,amssymb,natbib}
\usepackage{hyperref}
\usepackage{braket}
\usepackage{color}
\usepackage{float}
\usepackage{color,soul}

\begin{document}

\title{
Converting Faraday rotation into magnetization in europium chalcogenides
}

\author{S. C. P. van Kooten}
\altaffiliation[On leave from:]{Department of Applied Physics, Eindhoven University of Technology, Eindhoven 5612 AZ, The Netherlands}
\affiliation{Instituto de Fisica, Universidade de Sao Paulo, 05315-970 Sao Paulo, Brazil}
\author{P. A. Usachev}
\altaffiliation[Permanent address:]{Ioffe Institute, 194021 St. Petersburg, Russia}
\affiliation{Instituto de Fisica, Universidade de Sao Paulo, 05315-970 Sao Paulo, Brazil}
\author{X. Gratens}
\affiliation{Instituto de Fisica, Universidade de Sao Paulo, 05315-970 Sao Paulo, Brazil}
\author{A. R. Naupa}
\affiliation{Instituto de Fisica, Universidade de Sao Paulo, 05315-970 Sao Paulo, Brazil}
\author{V. A. Chitta}
\affiliation{Instituto de Fisica, Universidade de Sao Paulo, 05315-970 Sao Paulo, Brazil}
\author{G. Springholz}
\affiliation{Institut f\"ur Halbleiter und Festk\"orperphysik, Johannes Kepler Universit\"at Linz, 4040 Linz, Austria}
\author{A. B. Henriques}
\email{andreh@if.usp.br.}
\affiliation{Instituto de Fisica, Universidade de Sao Paulo, 05315-970 Sao Paulo, Brazil}
\begin{abstract}
We present a simple semiclassical model to sustain that in europium chalcogenides (EuX), Faraday rotation (FR) in the transparency gap is proportional to the magnetization of the sample, irrespective of the material's magnetic phase, temperature, or applied magnetic field.
The model is validated by FR and magnetization measurements in EuSe in the temperature interval 1.7--300~K, covering all EuSe magnetic phases (paramagnetic, antiferromagnetic type I or type II, ferrimagnetic and ferromagnetic).
Furthermore, by combining the semiclassical model with the explicit electronic energy structure of EuX, the proportionality coefficient
between magnetization and FR is shown to be dependent only on the wavelength and the band gap.
Due to its simplicity, the model has didactic value, moreover, it provides a working tool for converting FR into magnetization in EuX.
Possible extension of the model to other intrinsic magnetic semiconductors is discussed.
\end{abstract}
\date{\today}
\maketitle
\section{Introduction}
\label{sec:intro}
Faraday and Kerr rotations are powerful investigation tools of spin phenomena, the use of which have led, for example, to the demonstration of the spin Hall effect \cite{crooker,awschalomSQHE}.
Modern technology allows the measurement of extremely small Faraday rotation (FR) angles, in the nano radian range \cite{crookernrad}, so much that even the contribution of a single electron to the FR has been reported
\cite{badolato}.
Time-resolved FR gives access to fundamental parameters of spin coherence, such as
its formation and extinction times \cite{awschalomPT,bayerScience2006,renanPRB2015,kirilyukRMP,prl2018}.
However, in most reports FR is used only as an indicative measure of spin coherence, it is not converted numerically into the associated magnetization.
A quantitative connection between FR and magnetization is the subject of the present report.

In basic books on solid state physics \cite{madelung},
and in literature specialized on magneto-optics \cite{semiconductorsAndSemimetals,sugano,miuraBook},
FR per unit length of material is often described for diamagnetic materials, where FR is proportional to the magnetic field, $B$.
In a separate class of materials, the diluted magnetic semiconductors (DMS), FR was studied extensively, and various mechanisms of FR
have been identified \cite{GajKossut}.
However, the topic of the present investigation are concentrated, or intrinsic, magnetic semiconductors, whereby the magnetic atoms give origin to the top valence band, whose presence is essential for the FR, and therefore the mechanisms seen in DMS, or in diamagnetic semiconductors, do not apply.

For the diamagnetic case, a proportionality between FR and $B$ can be
justified by a simple classical model \cite{becquerel,sommerfeld}.
In contrast, in intrinsic magnetic semiconductors, the assumption of a constant FR/$B$ ratio fails squarely,
as shown in Ref.~\onlinecite{bauer}. In many magnetic semiconductors, FR is proportional to the magnetization, as demonstrated in Ref.~\onlinecite{mauger}, using Maxwell equations and the polarizability tensor for a cubic crystal. The proportionality between FR and magnetization has also been demonstrated
for EuTe \cite{prb17FR} and other concentrated
\cite{shenPR1964,shenbloembergenPR1964} and diluted magnetic semiconductors \cite{furdynaPRB1986},
using quantum mechanics, but these calculations are very involved, and require a detailed knowledge of the electronic structure of the investigated material. A simple conceptual model, based on elementary classical ideas, justifying that FR can be proportional to the magnetization in an intrinsic magnetic semiconductor, is lacking, and this work fills this gap.

In this paper we develop a simple semiclassical model, showing that in europium chalcogenides,
where the magnetic atoms are the source
of the highest energy valence band, for photon energies below the band gap, FR is proportional to the magnetization,
independently of the magnetic phase (paramagnetic, antiferromagnetic, ferrimagnetic,
or ferromagnetic), temperature, or magnetic field.
The model is validated by measurements of FR and magnetization in the 1.7-300 K temperature range and in 0-7~T magnetic fields.
The material chosen for the validation was the intrinsic magnetic semiconductor EuSe,
whereby by adjusting the external parameters all possible magnetic phases were covered. Our semiclassical model has the advantage over existing
quantum-mechanical theories due to its simplicity. Moreover, we show that for EuX, the proportionality constant between FR and magnetization is dependent only on the photon energy and the band gap of the
semiconductor.

The paper is organized as follows. In section \ref{sec:basics}, FR is introduced, in section \ref{sec:diamag}, the classical model for FR in a diamagnetic semiconductor is briefly reviewed, in section \ref{sec:mag} we introduce our semiclassical model for FR in EuX,
in section \ref{sec:test} the proportionality between FR and magnetization for EuSe in any scenario is thoroughly demonstrated experimentally, and in \ref{sec:VM} the semiclassical model is combined with the specific electronic energy structure of EuX, to obtain a working expression for the proportionality constant between magnetization and FR.

\color{black}
\section{Faraday rotation basics}
\label{sec:basics}
A linearly polarized light ray can be expressed as the superposition of two rays of equal intensity, one of which is
circularly polarized according to the right-hand rule (RCP), and the other according to left-hand one (LCP).
The superposition on these rays on exiting the sample gives the FR angle, per unit length, at the wavelength $\lambda$ \cite{maxwell,becquerel,semiconductorsAndSemimetals}:
\begin{equation}
\theta_F=\,\pi\, \frac{n_--n_+}{\lambda}
\label{eq:becquerel}
\end{equation}
where $n_\pm$ is the refractive index, and the plus or minus sign applies to RCP or LCP, respectively.
This formula shows that
circular birefringence, i.e. the inequality between $n_+$ and $n_-$, is the source of FR.

In general, semiconductor materials will contain several valence bands contributing to the birefringence.
Photons of energy within the band gap of the semiconductor are closest to resonance with the top valence band,
hence the polarization effects of lower lying bands can be discarded in a first examination.
The amplitude of the circular polarization
induced in the crystal by the
rotating electric field of the incoming light,
is given by \cite{kittel,feynman}
\begin{equation}
P_0^\pm=N \alpha^\pm E_0\
\label{eq:P0d}
\end{equation}
where $\alpha^\pm$ is the electronic polarizability of the atoms forming the valence band, $N$ is the number density of atoms in the solid,
and $E_0$ is the electric field amplitude of the RCP or LCP incident wave.
It should be emphasized that $\alpha$ in \eqref{eq:P0d} is the polarizability of an atom {\it embedded in the solid}, it is {\it not} the polarizability of an isolated atom. These polarizabilities are different from one another because the polarization of atoms, by light within the bandgap, is a perturbation and resonance effect.
The electron-photon interaction resonance depends on the spacing between electronic energy levels, which in the solid differ from that of the isolated atom,
due to energy band formation, hence the polarizability of an embedded atom differs from that of an isolated one.

On the other hand, taking the photoinduced polarization to be
the number density times the atomic polarizability, modified due to the atoms being embedded in the solid,
as by \eqref{eq:P0d}, is known to provide a very good description of the linear optical properties
of nonmetallic solids (see, for instance, Ref. \onlinecite{boyd},  section 1.4,  formula (1.4.16), which justifies the application of eq. \eqref{eq:P0d} to
describe the Faraday effect in europium chalcogenides.

Using the relation connecting the refractive index to the electronic polarizability \cite{feynman},
\begin{equation}
n_\pm^2=1+\frac{N}{\varepsilon_0}\alpha^\pm,
\label{eq:nOfAlpha}
\end{equation}
we arrive at
\begin{equation}
n_--n_+=\frac{n_-^2-n_+^2}{2n_0}=\frac{N}{\varepsilon_0}\frac{\alpha^--\alpha^+}{2n_0},
\label{eq:bir}
\end{equation}
where $n_0=(n_++n_-)/2$ is the refractive index that the material would have,
if no other valence band was present except the one under scrutiny, and $\varepsilon_0=8.85\times 10^{-12}$~F/m is the vacuum permittivity.

Equation \eqref{eq:bir} shows that for FR to be present, the induced polarization current
in the valence orbitals must be different for LCP and RCP. This is explored in the models below.

\section{Classical model of FR in a diamagnetic semiconductor}
\label{sec:diamag}
To build a clear contrast between FR in a magnetic semiconductor, i. e. one that contains atoms with unpaired electrons,
to FR in a diamagnetic semiconductor, whose electrons are all paired, let us review very briefly the well-known
classical model for FR in the diamagnetic semiconductor. This simple classical model is based on the Lorentz oscillator model of atoms,
and it is described in detail in Refs.~\onlinecite{becquerel,sommerfeld}.

For a non-magnetic atom, its angular momentum (orbital and spin) is zero, therefore there is no spatial orientation of the atom.
Hence by symmetry both RCP and LCP will induce equal and opposite polarization currents, therefore there will be no FR.
However, if a magnetic field is applied in the direction of light travel (i. e., in the Faraday geometry), then the Lorentz force will have opposite
effects on the LCP and RCP induced polarization currents, which become different from one another, and FR emerges. In the linear regime, the FR in this case is proportional to the intensity of the applied magnetic field.
\begin{figure}
\includegraphics[angle=0,width=90mm]{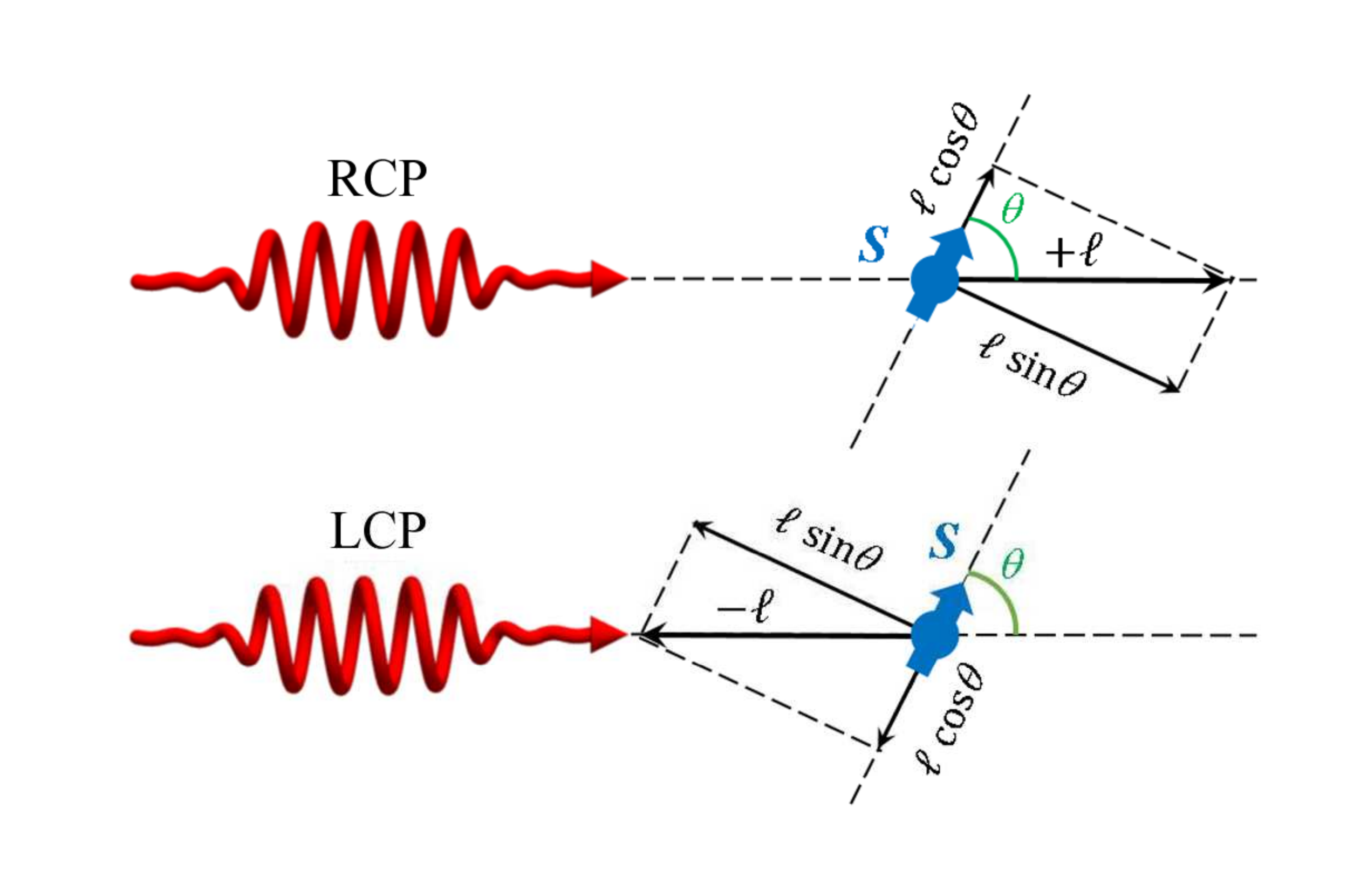}
\caption
{
The electric field $E$ of the linearly polarized light is equivalent to the superposition of RCP (top) and LCP light of equal amplitude (bottom),
carrying an angular momentum of $+\bm\ell$ and $-\bm\ell$, respectively.
}
\label{fig:LightOnAtom}
\end{figure}

\section{A semiclasical model connecting FR to the magnetization in
europium chalcogenides}
\label{sec:mag}

In EuX, the highest valence band states are formed by half-filled $4f$ orbitals of the Eu atom \cite{mauger}, which are buried deep within the ion, beneath the filled $5p$ shell,
hence the characteristics of the isolated orbital are well preserved \cite{blundellbook}, therefore the spin of the Eu atom, S=7/2, is maintained in the crystal.

In the semiclassical approach, the magnetic moment, or spin, of an atom,
is associated with a circulating electrical current,
whose direction and magnitude are described by a vector $\bm S$.
Let us inspect the interaction of the incident linearly polarized light with an average Eu atom in the solid, whose spin $\bm S$ makes an angle $\theta$ with the direction of light propagation, as depicted in figure \ref{fig:LightOnAtom}.
We express the incident light as a superposition of RCP and LCP,
which carry an angular momentum $+\bm\ell$ and $-\bm\ell$,
respectively, along the direction of light travel.
The angular momentum vector of the incoming RCP or LCP
light can be divided into
two components,
parallel and perpendicular to the vector $\bm S$,
as shown in figure \ref{fig:LightOnAtom}.
By symmetry, in a direction perpendicular to $\bm S$, RCP and LCP will induce identical polarizations in magnitude, but in opposite directions,
totaling zero.
Therefore birefringence must be associated with the circular polarization light induces parallel to $\bm S$, which is proportional to the projection of the light angular momenta onto $\bm S$. Hence when the angle between $\bm S$ and the direction of light travel is increased from zero to $\theta$, the induced polarization is  reduced by a factor of $\cos\theta$, i.e.
\begin{equation}
P_0^\pm= N \alpha^\pm_{||} E_0\,\cos\theta
\label{eq:P0}
\end{equation}
where $\alpha^\pm_{||}$ is the circular electronic polarizability of the solid,
when its spins are fully aligned with the direction of light travel ($\theta=0$ in figure \ref{fig:LightOnAtom}).

On the other hand, the magnetization projection in the direction of light propagation, $M$, is given by
\begin{equation}
M=N\mueff \cos\theta,
\label{eq:Mz}
\end{equation}
hence a comparison between equations \eqref{eq:P0} and \eqref{eq:Mz} leads to
\begin{equation}
P_0^\pm=\frac{M}{\mueff}\alpha^\pm_{||} E_0.
\label{eq:P01}
\end{equation}

Equating \eqref{eq:P01} and \eqref{eq:P0d} gives
\begin{equation}
\alpha^\pm=\frac{M}{\MSAT}\alpha^\pm_{||}
\label{eq:polariz}
\end{equation}
where $\MSAT=N\mueff$ is the saturation magnetization.

Substituting \eqref{eq:polariz} in \eqref{eq:bir}, and using \eqref{eq:becquerel}, we get
\begin{equation}
\theta_F^{\mbox{mag}}=\frac{\pi}{\lambda}\frac{N}{\varepsilon_0}\frac{M}{\MSAT}\frac{\alpha^-_{||}-\alpha^+_{||}}{2n_0}.
\label{eq:thetafmag}
\end{equation}

Equation \eqref{eq:thetafmag} shows that the contribution from Eu atoms to the FR
is proportional to their magnetization,
the proportionality coefficient being determined by the polarizability.
Because the polarizability is determined by the electronic energy structure,
the ratio $\theta_F/M$ will remain unchanged
as long as the relative position of the electronic energy levels, as well as their occupation, is invariant.
In a semiconductor, the essential parameter is the energy gap, $\EG$, between the valence and
the conduction bands.
If $k_BT\ll \EG$, the occupation of the electronic energy levels will be unchanged, which gives a measure of the range of temperatures in which $\theta_F/M$ is
expected to be constant in EuX, except for deviations due to band gap variations.
Thus $\theta_F/M$ behaves in the same fashion as the
refractive index of dielectrics, which is also tied
to variations of the band gap \cite{refrindexVsEG,refrIndexGaAs}.
As long as the photon energy is within the band gap, which is the situation considered in this work, contributions from other valence bands will generally be much smaller, due to their excitations being off-resonance with the incident photons, and the central result given by eq. \eqref{eq:thetafmag} will remain valid.

The significance and the value of the semiclassical model, with which the proportionality between FR and magnetization was demonstrated as expressed by eq. \eqref{eq:thetafmag}, based on a simple argument of forced oscillations and symmetry, can be well appreciated if we compare our model to the full quantum mechanical calculation, described in detail in Ref.~\onlinecite{prb17FR}, which requires the use of perturbation theory, Wigner rotations of spins, and statistical averaging. The end result is the same, but the semiclassical model is much simpler and transparent.

\section{Test of the semiclassical model in the magnetic semiconductor $\mbox{EuSe}$}
\label{sec:test}

\begin{figure}
\includegraphics[angle=0,width=90mm]{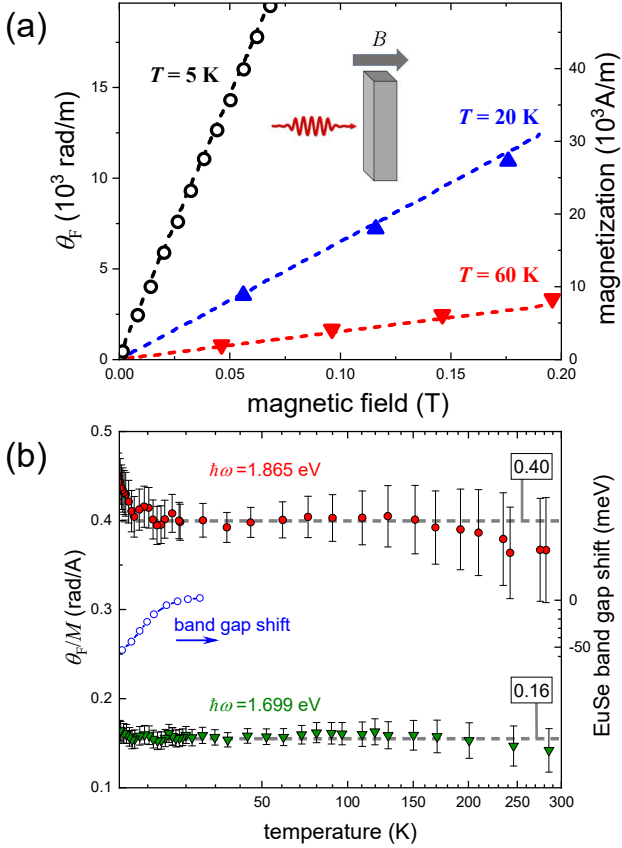}
\caption
{
(a) Lines depict FR, for photons of energy 1.865~eV, while dots represent magnetization, as a function of applied magnetic field, for $T$=5, 20 and 60~K. The magnetic field was applied perpendicular to the EuSe epitaxial layer;
(b) Ratio $\theta_F/M$ as a function of temperature for photons with energy 1.865~eV (full circles) and 1.699~eV (triangles).
The error bars were estimated at 15\% at low temperatures but increase towards room temperature, when the contribution to the epilayer becomes comparable to that of the substrate.
Below T=20~K, $\theta_F/M$ increases slightly for 1.865~eV, which is explained by the concomitant
narrowing of the band gap, shown by the empty circles, taken from Ref.~\onlinecite{wachter}.
}
\label{fig:ThetaFvsBandVvsT}
\end{figure}
\begin{figure}
\includegraphics[angle=0,width=90mm]{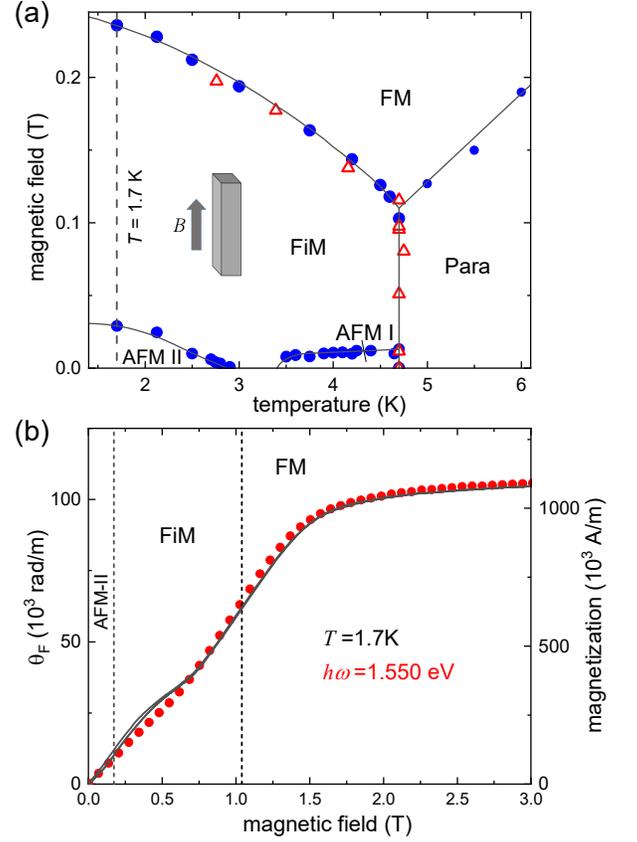}
\caption
{
(a) Magnetic phase diagram of the studied EuSe layer. The full and open dots represent data obtained from $M$ vs $B$ ($T=\mbox{const}$) and $M$ vs $T$ ($B=\mbox{const}$) traces, respectively. The magnetic field was applied parallel to the surface of the epitaxial sample. The solid lines are guides to the eye. The dotted line shows that at T = 1.7~K a magnetic field drives EuSe through an antiferromagnetic (AFMII), a ferrimagnetic (FiM), and a ferromagnetic (FM) phase.
(b) Magnetization (solid line) and FR at $\hbar\omega=1.55$~eV (dots), as a function of magnetic field, at $T=$1.7~K. The magnetic field was applied perpendicular to the surface of the epitaxial sample.  When B is applied perpendicular to the layer, the AFMII-FiM and FiM-FM phase boundaries are shifted to $B=0.17$~T and $B=1.04$~T, respectively, due to the demagnetization effect \cite{blundell}. Vertical lines show the boundaries between the magnetic phases indicated.
}
\label{fig:phasesAndThetaFvsM}
\end{figure}
\begin{figure}
\includegraphics[angle=0,width=90mm]{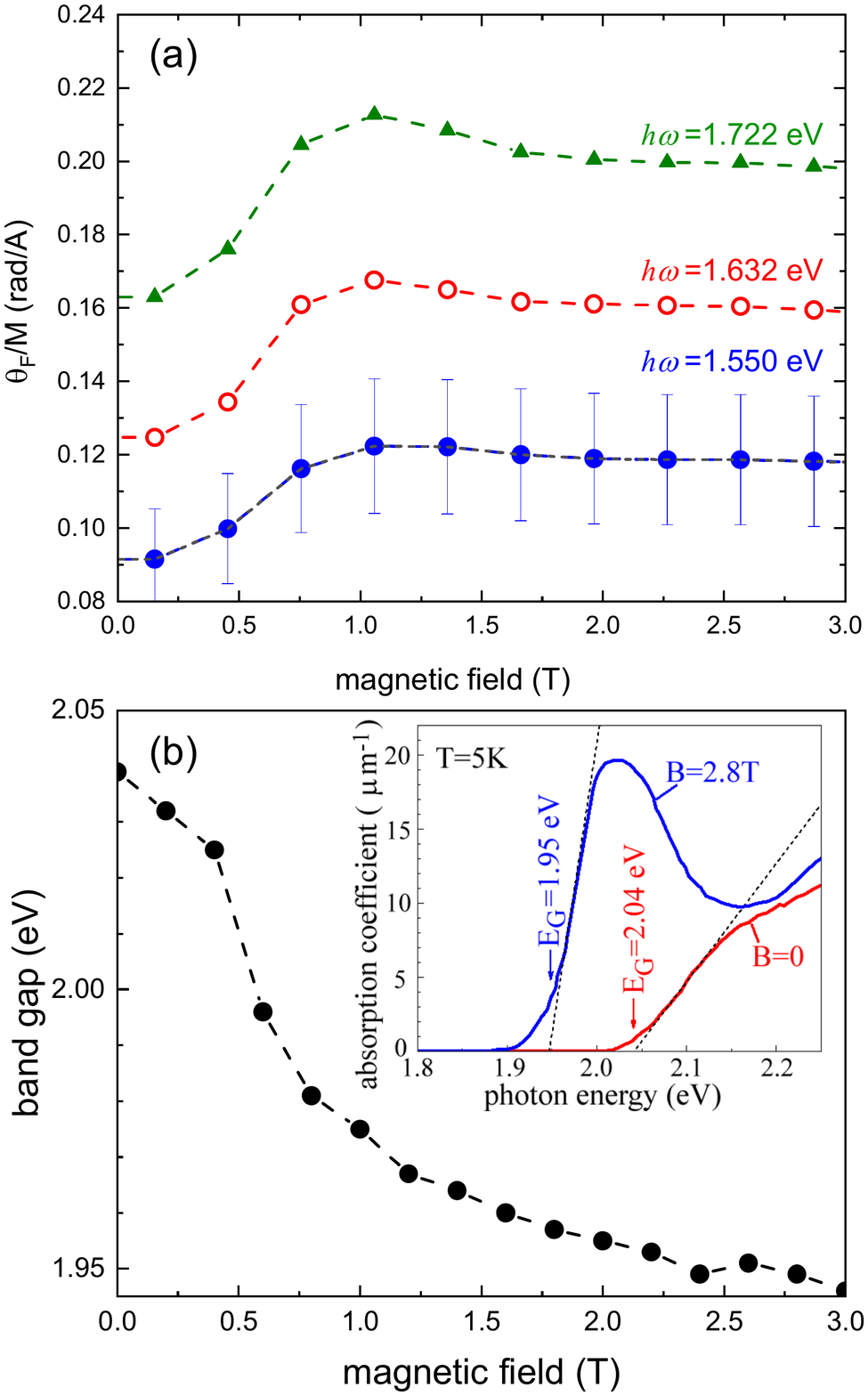}
\caption
{
(a) Ratio of the FR, at the indicated photon energies, to the magnetization, as a function of magnetic field, for $T=1.7$~K. At low fields, the ratio increases beyond the estimated uncertainty of 15\%; (b) EuSe band gap as a function of $B$, obtained from optical absorption measurements as shown in the inset. The absorption experiments are detailed in Ref.~\onlinecite{jpc07}.
}
\label{fig:belowTN}
\end{figure}

In the previous section it was argued that in EuX
FR is proportional to the magnetization. In this section this hypothesis is thoroughly tested using the magnetic semiconductor EuSe.
This material was chosen for the test because of its very rich magnetic phase diagram, therefore by applying a magnetic field and adjusting the temperature, the proportionality between FR and magnetization can be tested in all possible magnetic scenarios.

The EuSe crystalline samples were grown by molecular beam epitaxy
(MBE) onto (111) BaF$_2$ substrate.
Because of the almost
perfect lattice constant matching ($a=6.191~{\mbox{\AA}}$ and $a=6.196~{\mbox{\AA}}$ for EuSe and BaF$_2$, respectively), the
EuSe layer with $\mu$m thickness was bulklike and nearly
unstrained \cite{prb2008Diaz}. The data presented here was obtained on sample no. 1529, whose EuSe epilayer thickness is 2.5~$\mu$m.
The magnetization was measured using a SQUID magnetometer, which had a magnetic moment resolution better than 10$^{-11}$~Am$^2$.
The FR was measured using a linearly polarized beam from a semiconductor laser as the monochromatic light source,
and a polarization bridge containing balanced photodetectors. The contribution coming from the substrate to the FR was measured separately, using a substrate piece without the epilayer, and subtracted from the FR produced by the EuSe epilayer.

In fields up to 0.2~T at all temperatures, both the magnetization and the FR angle displayed a linear dependence
on $B$: typical results are shown in Figure~\ref{fig:ThetaFvsBandVvsT}(a).
It can be seen that the slopes of $\theta_F$ and $M$ vary several orders of magnitude with temperature,
however, the ratio $\theta_F/M$, obtained from the slopes for $B<0.2$~T,
\begin{equation}
\frac{\theta_F}{M}=\frac{d\theta_F/dB}{dM/dB},
\label{eq:ratio}
\end{equation}
and shown in figure~\ref{fig:ThetaFvsBandVvsT}(b) for the 4.8-300~K interval, remains constant.
The vertical bars represent the estimated experimental error, which increases towards room temperature, when the response from the substrate
becomes comparable to that of the epilayer, both in $\theta_F$ as well as in $M$ measurements.

For the photon energy of 1.865~eV, which is in near resonance with the band gap, $\thetaF/M$ increases slightly when the sample is cooled below 20~K.
This is explained by the concomitant narrowing of the gap (depicted
by the empty circles), which makes the light even closer to resonance with the band gap, which enhances the FR angle, as
explained at the end of this section.

In the temperature interval $4.8-300$~K examined so far, EuSe is in the paramagnetic phase, because its N\'eel temperature is $\TN=4.75$~K\cite{prl2018};
the magnetic phase diagram of our sample was measured, and it is shown in figure \ref{fig:phasesAndThetaFvsM}(a).
To investigate $\theta_F/M$ in the phases other than the paramagnetic one, the magnetization and FR, at various photon energies, were measured
at $T=1.7$~K as a function of field. At this temperature the magnetic field drives the EuSe sample through an antiferromagnetic, a ferrimagnetic, and a ferromagnetic phase, as figure \ref{fig:phasesAndThetaFvsM}(a) shows (see also \cite{springholzprl}).
A comparison of the magnetization and FR curves can be seen in figure~\ref{fig:phasesAndThetaFvsM}(b), both exhibit an almost identical dependence on $B$,
minor differences in the $B$-dependencies are within the range of the experimental uncertainties.
The dependence of $\theta_F/M$ on $B$
is shown in fig.~\ref{fig:belowTN}(a), for various photon energies. At low fields, $\theta_F/M$ remains at the same value measured in the paramagnetic phase up to 300~K.

Increasing the field, $\theta_F/M$ increases and tends to a saturation. This process can be understood within the frame of the semiclassical model of sections \ref{sec:basics} and \ref{sec:mag}.
When $B$ is increased at T=5K, the EuSe bandgap narrows, as shown in figure~\ref{fig:belowTN}(b), because the applied field imposes ferromagnetic order over a paramagnetic lattice, which lowers the energy of the conduction electrons due to the band-lattice exchange interaction.
The dichroic spectrum showing the LCP/RCP splitting was investigated in Ref.~\onlinecite{prb05}.
A similarly large bandgap redshift of about 100~meV by applying a magnetic field is also observed in YIG \cite{aplYIG}, and it is also associated with the conduction band-lattice exchange interaction. As figure~\ref{fig:belowTN} shows, the bandgap redshift saturates around 2.5~T
\footnote{The band gap was determined by extrapolating the linear region in the absorption curve to the abscissa,
as shown in the inset of figure~\ref{fig:belowTN}(b). Other authors have estimated the band gap through the photon energy at which the optical transmittance falls to 1\% of the transparency region \cite{heiss}, but this leads to a band gap dependent on the thickness of the sample; the band gap is commonly estimated by a Tauc plot \cite{tauc}, but this does not apply here, because a Tauc plot relies on the on parabolic dispersion, which is not the case for EuX, where the 5$d(t_{2g})$ conduction band has a tight-binding character \cite{maugerTB,prb09}.}.
The increase of the FR angle with increasing applied magnetic field is because the photon energy of the incident light becomes closer to resonance with the bandgap, which implies that the
electronic polarizability increases, as the classical Lorentz model of forced atomic oscillators predicts.

\section{Proportionality constant between magnetization and FR in europium chalcogenides}
\label{sec:VM}
\begin{figure}
\includegraphics[angle=0,width=86mm]{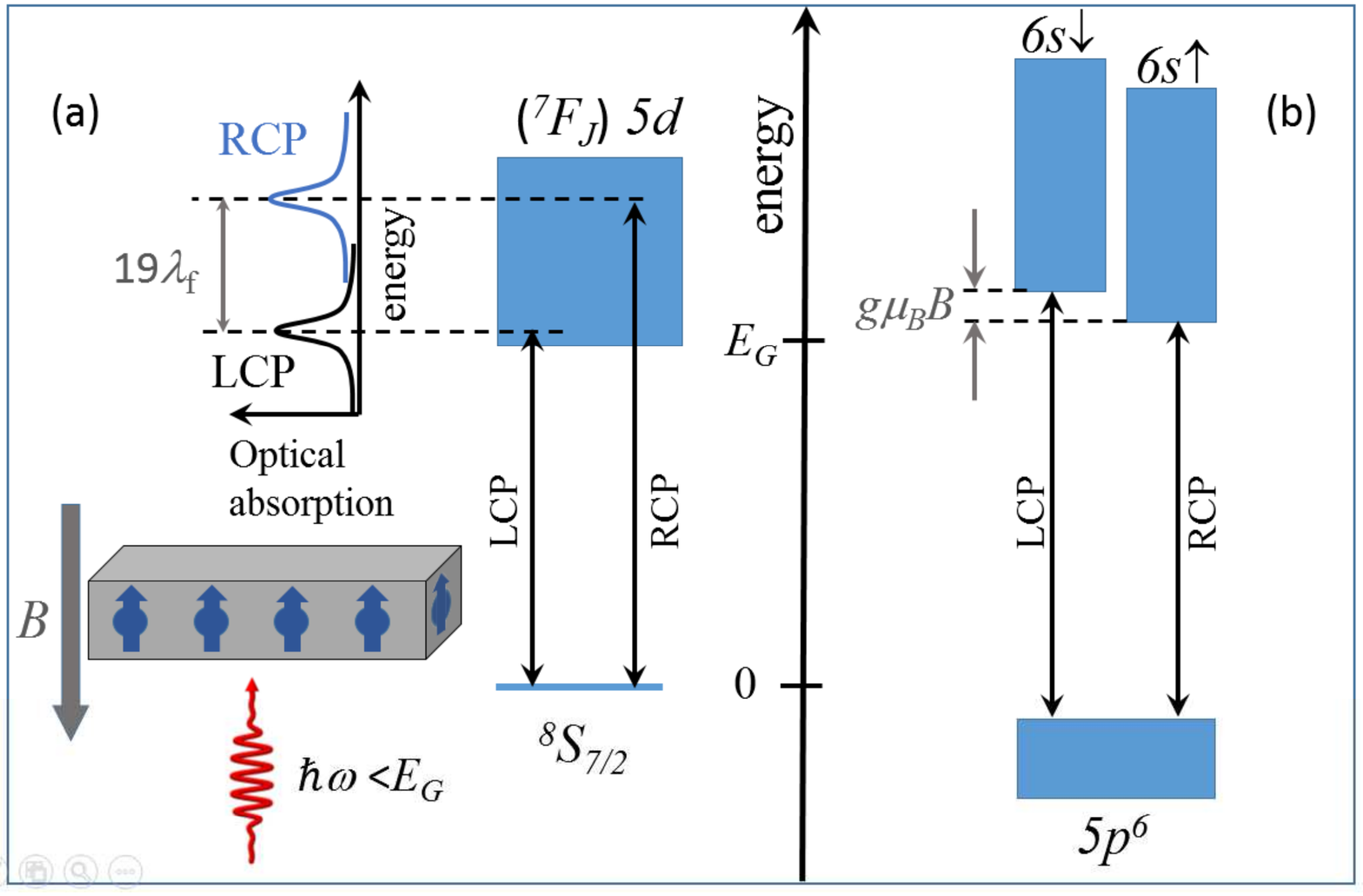}
\caption
{
Electronic levels in EuTe. (a) Under a strong magnetic field, the Eu spins are aligned ferromagnetically, and the
absorption spectrum shows a strong magnetic circular dichroism, whereby the RCP and LCP absorption peaks, corresponding to transitions from the Eu atoms in the $S_{7/2}$ state
to the $5d(t_{2g})$ conduction band, are split by $\sim 19 \lambda_f$, where $\lambda_f$ is the spin-orbit coupling constant for the Eu$^{3+}$ ion \cite{prb05}
(b) RCP and LCP optical transitions between the valence band formed by $5p$ orbitals of the Te atoms and a conduction band formed by $6s$ Eu states \cite{mauger,wachter}.
}
\label{fig:EuTe}
\end{figure}

In section \ref{sec:mag} we showed that in EuX FR is proportional to the magnetization.
Here we go a step forward, and determine the proportionality coefficient in EuX, using their well-known specific electronic energy structure, sketched in figure~\ref{fig:EuTe} \cite{mauger}.
First we shall estimate the weight of the various valence bands to the FR.
Using quantum mechanical time dependent perturbation theory \cite{boyd} applied to EuTe it can be shown that
\footnote{In arriving at \eqref{eq:alphaQM}, we substituted $\alpha^\pm=\varepsilon_o\chi^\pm/N$ in equation (9) of Ref.~\onlinecite{prb17FR}, where $\chi$ is the susceptibility.}
\begin{equation}
\alpha^\pm=\frac{1}{2}\sum_n \frac{\left|\mu^\pm_{gn}\right|^2}{E_{ng}-\hbar\omega},
\label{eq:alphaQM}
\end{equation}
where $\ket{g}$ represents the ground state of energy $E_g$, described by an electron in the valence band,
$\ket{n}$ represents the excited states of energy $E_n$, corresponding to an electron in the conduction band, $E_{ng}=E_n-E_g$,
and $\mu_{gn}^\pm$ is the electric dipole matrix element
\begin{equation}
\mu^{\pm}_{gn}=-e\bra{n}x\pm i y\ket{g}.
\label{eq:matel}
\end{equation}
Equation \eqref{eq:alphaQM} shows that the coupling through light between the ground and excited electronic states
determines the induced polarization current.

In EuTe, when the spins are aligned with the direction of light by the application of a large magnetic field,
the band-edge optical absorption becomes strongly dichroic \cite{jpc04,prb05,jpc07,prb08},
and shows two narrow peaks, one for LCP and another for RCP,
split by $\sim 19\lambda_f$ \cite{prb05}, where $\lambda_f$ is the spin-orbit coupling constant
for the Eu$^{3+}$ atom, as sketched on the left of figure \ref{fig:EuTe}.
The circular dichroism is associated with optical transitions between the valence level, $^{8}S_{7/2}$, formed by the magnetic Eu atoms, into the $5d(t_{2g})$ conduction band.
The equal height of the two lines implies that $\left|\mu^+_{gn}\right|^2\sim \left|\mu^-_{gn}\right|^2\sim\mu_{df}^2$.
Taking the approximated absorption spectrum into account, \eqref{eq:alphaQM} leads to
\begin{equation}
\alpha^-_{||}\sim\frac{1}{2}\frac{\mu_{df}^2}{\EG-\hbar\omega},\,\,\,\alpha^+_{||}\sim\frac{1}{2}\frac{\mu_{df}^2}{\EG+19\lambda_f-\hbar\omega}.
\label{eq:alphaEuTe}
\end{equation}
Substituting \eqref{eq:alphaEuTe} in \eqref{eq:thetafmag}, and using \eqref{eq:nOfAlpha}, we get
\begin{equation}
\theta_F^{\mbox{\small mag}}\sim\frac{\pi}{\lambda}\frac{M}{\MSAT}\frac{n_0^2-1}{2n_0}\frac{19\lambda_f}{\EG-\hbar\omega}.
\label{eq:FMag}
\end{equation}
In obtaining \eqref{eq:FMag}, the condition $19\lambda_f\ll \EG-\hbar\omega$ was assumed, which requires the incoming photons to be sufficiently away from resonance with the band gap.

It must be emphasized, however, that to arrive at \eqref{eq:FMag}, we \emph{did not} make any approximation concerning the EuX band edge electronic energy structure, we \emph{did not} substitute the energy bands of EuTe by
zero width atomic energy levels. We worked within the frame of the full ($^{8}S_{7/2}$-valence band, $5d(t_{2g})$-conduction band) model, which is specific for EuX, and which has successfully described such effects as the continuous evolution of a 500~meV broad featureless absorption band at zero field to a doublet of sharp dichroic lines in high fields \cite{jpc04,prb05,jpc07,prb08}, as well as second harmonic generation \cite{prl09,prb09}, linear dichroism \cite{jpc08}, and Faraday rotation \cite{prb17,prb17FR} in EuTe and EuSe. We simply exploited the well-known experimental fact that at high fields, when all Eu spins align ferromagnetically, two birrefringent narrow lines emerge in the optical absorption threshold of EuX, their width being much less than the band gap \cite{jpc04,prb05,prb08}, which has nothing to do with a substitution of bands by energy levels.

Now we shall inspect the contribution to the FR coming from
valence bands generated by completely filled atomic states, whose magnetic moment is zero.
In EuTe, just below the $^8S_{7/2}$ valence state, there is a valence band built from $5p^6$ shells of the Te atoms \cite{mauger}, which can be polarized by the incoming light through the
dipole-allowed admixture of empty $6s$ states,
as indicated on the right-hand side in figure \ref{fig:EuTe}. In the absence of a magnetic field, the $5p$-band will give no birefringence,
because optical absorption associated with $5p\rightarrow 6s$ transitions will be identical in position and strength for RCP and LCP. But when
a magnetic field is applied, the RCP and LCP absorption bands are split by the Zeeman energy, $g\mu_B B$. Then, proceeding exactly in the same way as when obtaining \eqref{eq:FMag},
the diamagnetic contribution to the FR is found to be
\begin{equation}
\theta_F^{\mbox{\small diamag}}\sim-\frac{\pi}{\lambda}\frac{n_1^2-1}{4n_1}\frac{g\mu_B B}{\EG-\hbar\omega},
\label{eq:Fd}
\end{equation}
where $n_1$ is the refractive index associated with the diamagnetic valence band. Equation~\ref{eq:Fd}
shows that the diamagnetic FR is proportional to the magnetic field, in stark contrast to the contribution from the magnetic valence band,
given by eq.~\eqref{eq:FMag}, which is proportional
to the magnetization.

Dividing \eqref{eq:FMag} by \eqref{eq:Fd},
the relative weight of the diamagnetic FR is found
\begin{equation}
\left|\frac{\theta_F^{\mbox{\small diamag}}}{\theta_F^{\mbox{\small mag}}}\right|\sim \frac{\MSAT}{M}\,\,\frac{g\mu_B B}{19\lambda_f},
\label{eq:thetaMoverThetad}
\end{equation}
where $n_0\sim n_1$ was used.
Given that $g\mu_B=0.116$~meV/T, and that $19\lambda_f=180$~meV \cite{prb05},
then the diamagnetic FR
will generally be much smaller than the magnetic one.

\begin{figure}
\includegraphics[angle=0,width=90mm]{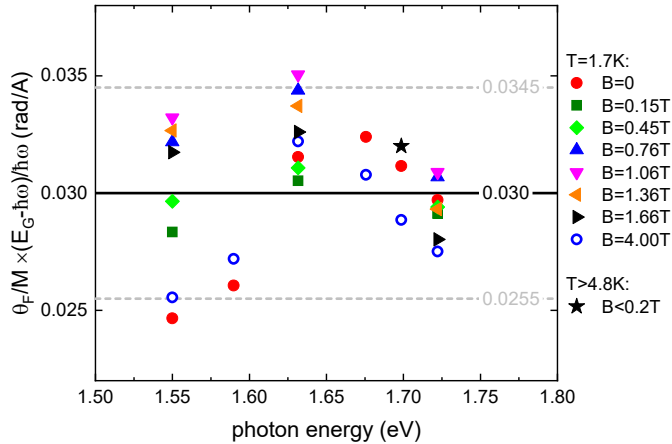}
\caption
{
Measured ratio $\theta_F/M$ multiplied by $(E_G-\hbar\omega)/\hbar\omega$, as a function
of photon energy, for various magnetic fields at T=1.7K,
and in 1.7-300~K interval, for fields $B\leq 0.2$~T. The thickness of the EuSe epilayer was 2.5~$\mu$m.
The average is shown by the full line, and a 15\% deviation is shown by the dashed lines.
}
\label{fig:Const}
\end{figure}

We shall take a step further, to investigate in more detail the proportionality coefficient between magnetization and FR. We rewrite equation \eqref{eq:FMag} as
\begin{equation}
\frac{\theta_F}{M}\frac{E_G-\hbar\omega}{\hbar\omega}={\mbox{const}},
\label{eq:const}
\end{equation}
where the constant is determined by the materials refractive index, spin-orbit coupling constant, and saturation magnetization.
Equation \eqref{eq:const} was tested by plotting $\theta_F/M$ as a function of photon energy, using data taken in a wide temperature and magnetic
field range, covering all EuSe magnetic phases,
and figure \ref{fig:Const} shows the result.
It can be seen that the data points deviate from the average of 0.03~rad/A by at most 15\%, which is
the estimated error bar in our experiment, also shown  in figure~\ref{fig:Const}. The deviation at lower energies is larger, because the FR angle is smaller, hence the experimental error is larger.
This is quite remarkable, because the data shown on figure~\ref{fig:Const} covers a 300~K temperature and a 0-7 magnetic field interval, respectively, where all possible magnetic phases occur, and where the ratio $\theta_F/M$ changes by an order of magnitude, and where $M$ and $\theta_F$ vary several orders of magnitude, nevertheless (17) remains constant within experimental error.
This not only validates our semiclassical model, it makes equation \eqref{eq:const} a practical  formula
to describe $\theta_F/M$ in all circumstances, substituting a full complex quantum-mechanical calculation in EuX.

Equation~\eqref{eq:const} should remain valid for other intrinsic magnetic semiconductors, in which a valence level is formed by strongly localized atomic orbitals with non-zero
magnetic moment. For the case of magnetic semiconductors where the top valence band is diamagnetic, FR will be a superposition
of one component proportional to the magnetization, and another proportional to $B$.
As an example, GdN  has a top valence band built from nitrogen
$2p$ states \cite{prb2013_GdN}, situated above the localized $^8S_{7/2}$ valence state of the Gd rare earth atoms (i.e., the position of the $p$ valence band and of the localized $^8S_{7/2}$ valence level, shown in figure~\ref{fig:EuTe} for EuTe and EuSe, are inverted in order). The magnetic circular dichroism observed in GdN \cite{aplGdN} is an indication that FR will also be connected to the magnetization in this system. However, because in GdN the $p$ valence band forms the band gap, for below-the-gap FR the diamagnetic contribution may be of the same magnitude as the ferromagnetic one, making a more complicated scenario. An analysis of the relationship between FR and magnetization in GdN and its dependence on the photon energy will require a separate investigation.

\section{Conclusion}
We developed a semiclassical model to show that in europium chalcogenides
the FR is proportional to the magnetization. The model is based on classical physics concepts only. Our model for FR in the magnetic semiconductors EuX adds to the well known classical model of FR in a diamagnetic semiconductor, forming a didactic picture of the diversity of the FR in different solids. The model is validated by data taken on EuSe in a large temperature and magnetic field range, covering all possible magnetic phases.
Moreover, we provided a formula connecting the Faraday rotation angle, the magnetization, the photon energy and the semiconductor band gap, which is a valuable practical solution for the conversion of FR into magnetization, at any temperature
and magnetic phase in any member of the  EuX family.

\section{Acknowledgements}
This work was funded by
CNPq (Projects 307400/2014-0 and 456188/2014-2) and
FAPESP (Project 2016/24125-5). P. A. U. acknowledges support by the Russian Science
Foundation (Project 17-12-01314). G. S. acknowledges support by the Austrian Science Funds, Project P30960-N27.

\bibliographystyle{apsrev4-1}

%

\end{document}